# Development of a Laser-based angle-resolved-photoemission spectrometer with sub-micrometer spatial resolution and high-efficiency spin detection

R. Z. Xu[1,2,*], X. Gu[1,2,*], W. X. Zhao[1,2,*], J. S. Zhou[1,2], Q. Q. Zhang[1,2], X. Du[1,2], Y. D. Li[1,2], Y. H. Mao[3], D. Zhao[4], K. Huang[4], C. F. Zhang[3], F. Wang[6], Z. K. Liu[5,6], Y. L. Chen[5,6,7‡] and L. X. Yang[1,2‡]

[1]State Key Laboratory of Low Dimensional Quantum Physics, Department of Physics, Tsinghua University, Beijing 100084, China.
[2]Frontier Science Center for Quantum Information, Beijing 100084, China.
[3]College of Advanced Interdisciplinary Studies, National University of Defense Technology, Changsha, Hunan 410073, China.
[4]Department of Optics and Optical Engineering, University of Science and Technology of China, Hefei, Anhui 230026, China.
[5]School of Physical Science and Technology, ShanghaiTech University and CAS-Shanghai Science Research Center, Shanghai 201210, China.
[6]ShanghaiTech Laboratory for Topological Physics, Shanghai 200031, China.
[7]Department of Physics, Clarendon Laboratory, University of Oxford, Parks Road, Oxford OX1 3PU, UK.
[*]These authors contribute to this work equally.
[‡]Email address: LXY: lxyang@tsinghua.edu.cn, YLC: yulin.chen@physics.ox.ac.uk

**Angle-resolved photoemission spectroscopy with sub-micrometer spatial resolution (μ-ARPES), has become a powerful tool for studying quantum materials. To achieve sub-micrometer or even nanometer-scale spatial resolution, it is important to focus the incident light beam (usually from the synchrotron radiation) using X-ray optics such as the zone plate or ellipsoidal capillary mirrors. Recently, we developed a laser-based μ-ARPES with spin-resolution (LMS-ARPES). The 177 nm laser beam is achieved by frequency doubling a 355 nm beam using a KBBF crystal and subsequently focused using an optical lens with a focal length of about 16 mm. By characterizing the focused spot size using different methods and performing spatial-scanning photoemission measurement, we confirm the sub-micron spatial resolution of the system. Compared with the μ-ARPES facilities based on the synchrotron radiation, our LMS-ARPES system is not only more economical and convenient, but also with higher photon flux (> $10^{14}$ photons/s), thus enabling the very high resolution and statistics measurements. Moreover, the system is equipped with a two-dimensional spin detector based on exchange scattering at a surface-passivated iron film grown on a W(100) substrate. We investigate the spin structure of the prototype topological insulator $Bi_2Se_3$ and reveal high**

**spin-polarization rate, confirming its spin-momentum locking property. This lab-based LMS-ARPES will be a powerful research tool in studying the fine electronic structures of different condensed matter systems, including topological quantum materials, mesoscopic materials and structures, and phase-separated materials.**

## I. INTRODUCTION

The investigation of the electronic structure of quantum materials not only provides fundamentally important scientific implication but also lays the foundation for the exploration and application of new materials [1]. With the capability of measuring the energy and momentum of electrons, angle-resolved photoemission spectroscopy (ARPES) can directly visualize the electronic structure of crystalline materials with crucial information including band dispersion, Fermi surface topology, band gap, carrier density, and electron effective mass, etc. Moreover, it can extract subtle information such as electron-electron correlation, electron-phonon interaction, and spin-orbit coupling [2, 3]. In the past decades, ARPES has become a powerful tool for the exploration of numerous intriguing quantum phenomena such as high-temperature superconductivity [4-7], heavy-fermion materials [5, 8, 9], charge/spin density waves [10-12], magnetism [13, 14], and topological quantum physics [15-17].

With the rapid development of condensed matter physics, richer details of electronic structures are desired, including spatial variation/distribution, spin polarization, and non-equilibrium dynamics, which urges the improvement of the state-of-the-art ARPES with new probe capability [2, 15, 18-27]. Recently, by focusing the incident light to an extremely small spot, μ-ARPES with high spatial resolution has exhibited its power in studying the local electronic structure of materials with nano- or micro-meter scale structures, such as the phase-separated iron-based superconductor [28, 29], two-dimensional materials [30, 31], and twisted ultrathin films [32, 33]. It also promises direct detection of topological edge states [34]. The key to enhance the spatial resolution is to focus the light beam, which demands coherent light source. As a result, usually the spatially-resolved nano- or µ-ARPES systems are constructed based on synchrotron radiation [23, 35-39]. Because of the extremely short beam wavelength, the beam spot can be focused to smaller than 100 nm using Fresnel zone plate [35, 36]. However, the synchrotron radiation facility is very expensive, which limits the broad application of nano- and µ-ARPES. Besides, the optical transmissivity of the zone plate is usually low, which reduces the experimental efficiency as well as the energy resolution since larger analyzer slit and higher pass energy are needed to improve the data statistics. By contrast, laser source with high intensity provides another possibility to achieve high spatial resolution. Using $\beta$-$BaB_2O_4$ (BBO) or $KBe_2BO_3F_2$ (KBBF) crystals, the laser of energy up to 6 or 7 eV can be delivered with high coherence and photon flux. The laser beam can be easily focused to micro-meter scale using either zone plate or optical lens [40], which is an ideal photon source for lab-based µ-ARPES at low cost.

On the other hand, as an intrinsic property of electrons, the spin degree of freedom plays an important role in many emergent phenomena such as magnetism, topological quantum physics, superconductivity, and so on. Extensive research efforts have been made to achieve efficient measurement of spin-polarized electronic structure [41-45]. In principle, a spin-detector can be built based on either Mott scattering at high-$Z$ material targets such as gold [46] and iridium [47] or exchange scattering at ferromagnetic targets such as Fe(001) surface [48]. The detection efficiency of the latter can be 20 times higher than the former, while the lifetime of the former is much longer than the latter. Very recently, it is proposed that the lifetime of Fe(001) target can be significantly prolonged from several hours to weeks, with the high detection efficiency maintained when covered with monolayer $p(1\times1)$ ordered oxygen [49], which provides an ideal scattering target for spin detection. To further enhance the spin detection statistics, high-photon flux is desirable, which can be easily fulfilled by laser source as well.

In this paper, we report the development of multifunctional laser-based µ- and spin-resolved ARPES (LMS-ARPES) system combining sub-micron spatial resolution and high-efficiency spin detection. The 177 nm deep ultraviolet (DUV) laser is delivered by frequency doubling of fundamental 355 nm commercial laser in a KBBF crystal. We achieve sub-micron laser beam size by focusing the DUV laser using a commercial lens with focal length of 16 mm. By scanning the beam spot across a knife edge, we determine the beam size at focal point to be smaller than $1 \times 1$ μm$^2$. We also perform the scanning photoemission measurement on a 1 μm structured gold pattern, which exhibits sub-micron spatial resolution of our laser-based μ-ARPES. The system is also equipped with a two-dimensional spin detector based on very low energy electron diffraction (VLEED) on a surface-passivated Fe/W(100) target. We successfully reveal the spin polarization of the topological surface states (TSSs) of the prototypical topological insulator Bi$_2$Se$_3$, confirming its spin-momentum locking property. The spin polarization rate of the TSSs is as high as 84%, suggesting high spin detection efficiency of the system. With both sub-micron spatial resolution and high-efficiency spin detection, our LMS-ARPES system features great potential for exploring many important frontiers of condensed matter physics.

## II. OVERVIEW OF THE LMS-ARPES

Figure 1(a) schematically illustrates the basic working principle of laser-based μ-ARPES. The 177 nm laser is focused using an optical lens and the photoelectrons are collected with a semi-sphere analyzer. The advantage of the laser source and focusing lens over the synchrotron radiation and zone plate is the low cost and high photon flux. Figure 1(b) shows the overview of our LMS-ARPES system that mainly includes four parts: I. Micro-focused 177 nm laser system; II. Sample manipulator; III. Analysis Chamber, vacuum system, and electron analyser (ScientaOmicron DA30L); IV. VLEED spin detector.

Figure 1(c) details the light path of the laser source. A commercial solid-state laser (Paladin-4000, Coherent. Inc.) provides 355 nm laser beam [blue beam in Figs. 1(b) and (c)] which is attenuated and focused in a KBBF crystal to generate 177 nm DUV laser. The output DUV laser propagates in an angle of about 9° from the fundamental beam. Since the DUV laser is not in ideal TEM$_{00}$ mode, it exhibits anisotropic divergence and highly elliptical beam spot. Therefore, we use two cylindrical lenses to reshape the beam spot. To minimize the focused spot size, we insert a 1:10 beam expander in the light path before focusing the beam onto the sample. A wave plate is inserted to adjust the polarization of the beam. In order to reduce the absorption of the photons by the air, the DUV optics are all sealed in a chamber filled with high-purity nitrogen. Finally, the expanded 177 nm laser beam with diameter of about 1 cm is guided into the analysis chamber through a CaF$_2$ window and focused on the sample for ARPES measurement. The final focusing can be achieved using either a zone plate or a lens. Here we choose a commercial lens with a focal length of 16 mm for high transmission and to avoid the side spots caused by the zone plate.

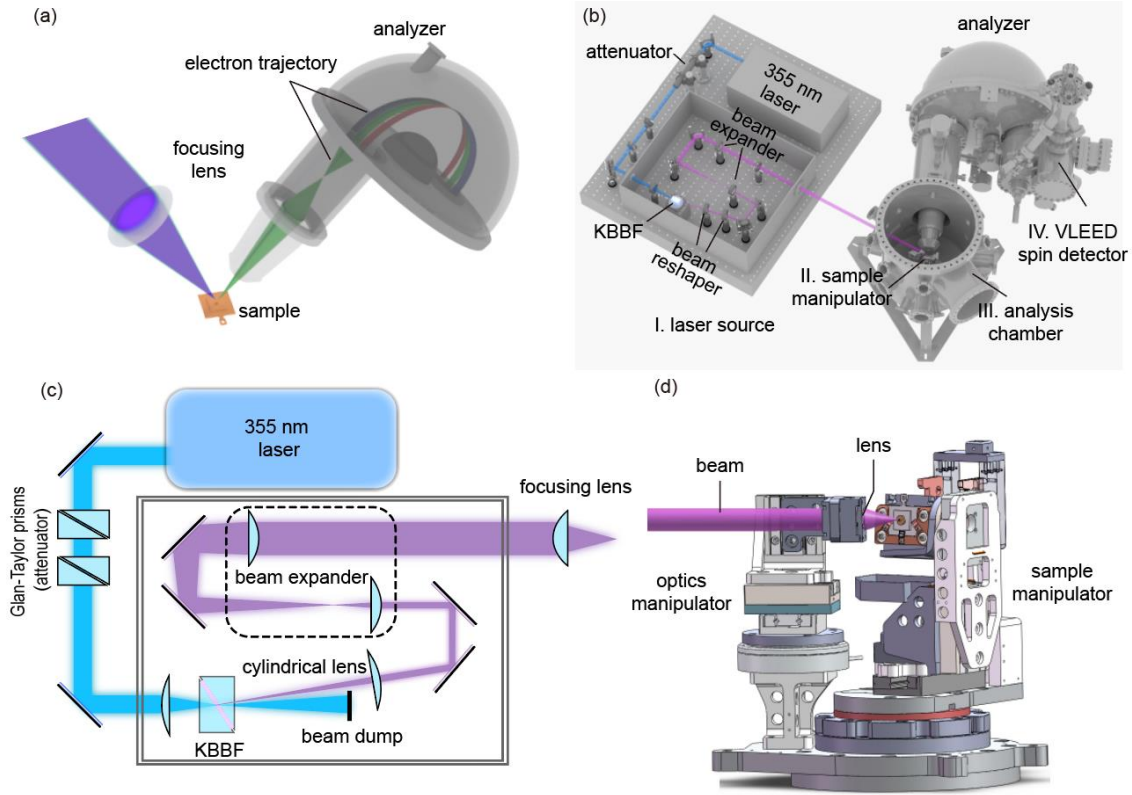

FIG. 1. Overview of LMS-ARPES systems. (a) Schematic illustration of the working principle of laser-based ARPES. (b) Layout of the LMS-ARPES. (c) Details of the beam path of the laser source. (d) Illustration of the sample and lens manipulators.

To guarantee a high-quality µ-ARPES measurement, the sample manipulator needs to provide sufficiently accurate control of the sample degrees of freedom. We adopt a self-assembled 5-axes manipulator using piezo stages (®Smaract) to control the sample position and angles, as shown in Fig. 1(d). The positions ($x$, $y$ and $z$) and the angles (polar and tilt) are accurately controlled by translational and rotary piezo stages, respectively. The translation and rotation accuracies are 10 nm and 25 µ° respectively, which are sufficiently high for µ-ARPES measurements. To conduct temperature-dependent measurements, we connect the thermally isolated sample stage to a commercial open-cycle cold stage with a copper braid. The sample temperature can be adjusted in a large range from 14 K to 380 K. The technical parameters are summarized in Table 1.

Table 1. The technical parameters of the LMS-ARPES system.

| Parameters | Values |
| --- | --- |
| Photon Energy | 7 eV |
| Photon Flux (after focus) | $> 10^{14}$ photons/s |
| Pressure | $\leqslant 1 \times 10^{-10}$ mbar |
| Spatial Resolution | $\leqslant 1$ µm |
| Sample Temperature | 14.5 K ~ 380 K |
| Sample Degrees of Freedom | 5 |
| Manipulation Precision | 10 nm for translation<br>25 µ° for rotation |
| Spin Reflection | $> 20\%$ @ SE = 8 eV |

## III. SPATIAL RESOLUTION OF LMS-ARPES

Since the beam is collimated before entering the UHV chamber, we can test the focusing result in the nitrogen-filled chamber using the same focusing lens. We perform two measurements to characterize the focused spot size, as illustrated in Fig. 2. We first scan the beam across a knife-edge and use a phototube to collect the unblocked photons [Fig. 2(a)]. The beam size is determined by the spatial distribution of photon flux. As shown in Figs. 2(b) and (c), the signal shows a step whose width is determined by the convolution of the beam diameter and the knife edge. The fit of the derivative of the signal to a Gaussian suggests a full width at half maximum (FWHM) of about 1.0 μm and 0.94 μm along the horizontal and vertical directions respectively. Considering the diffraction effect [50] and the sharpness of the knife edge, the actual beam size should be smaller than $1 \times 1$ μm$^2$.

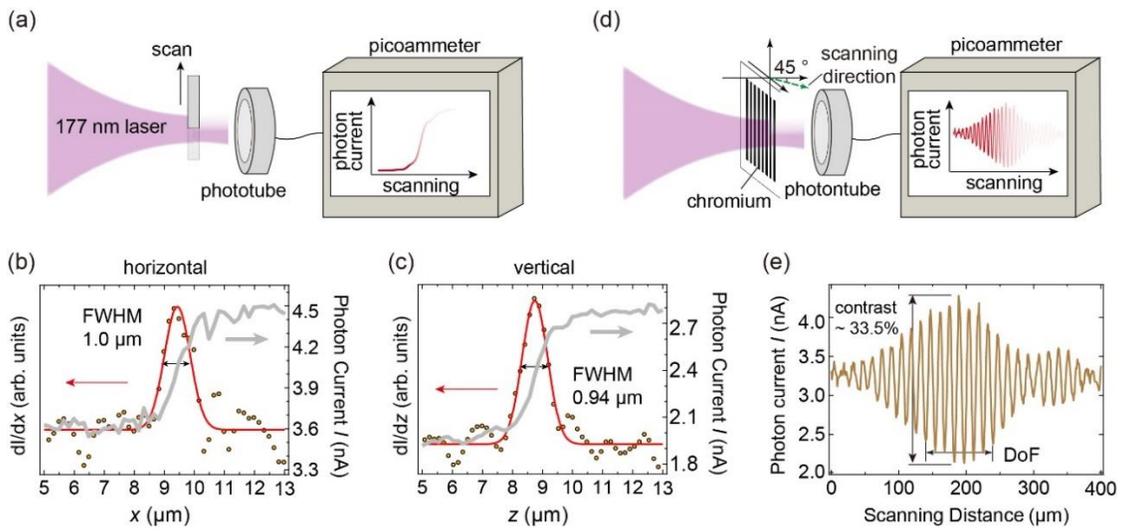

FIG. 2. Characterization of the beam size focused by the commercial lens. (a) Schematic illustration of the measurement by beam scanning across a knife-edge. (b, c) Photon current (grey curves) as a function of scanning distance along horizontal (b) and vertical (c) directions. The black circles show the derivative of the photon current curve with respect to the scanning distance and the red curves are the fits to a Gaussian (red curves). (d) Schematic illustration of the diagonal scanning measurement (see the main text for details). (e) The photon current after transmission through the 1 μm pattern as a function of scanning distance.

To double check the beam size, we perform a "diagonal focusing scan" of the beam across a micro-structured periodic pattern with 2 μm periodicity [Fig. 2(d)], which is made of chromium deposited on the CaF$_2$ substrate. Therefore, the beam will either be blocked by the chromium stripes or transmit through the CaF$_2$ substrate. The pattern is translated along a direction at 45° with respect to the light path, which ensures that, on the one hand, the beam scans across the pattern to examine spatial resolution; and on the other hand, the pattern scans along the beam path so that the focal point can be reached, where the contrast of the transmission intensity is highest. Figure 2(d) shows the measured photon current as a function of scanning distance. The highest contrast at the focal point is about 33%, suggesting that the beam spot is sufficiently small to resolve 1-micron structures. Moreover, the diagonal focusing scan also suggests that the focusing depth of field is about 50 μm size, which is ideal for μ-ARPES measurements.

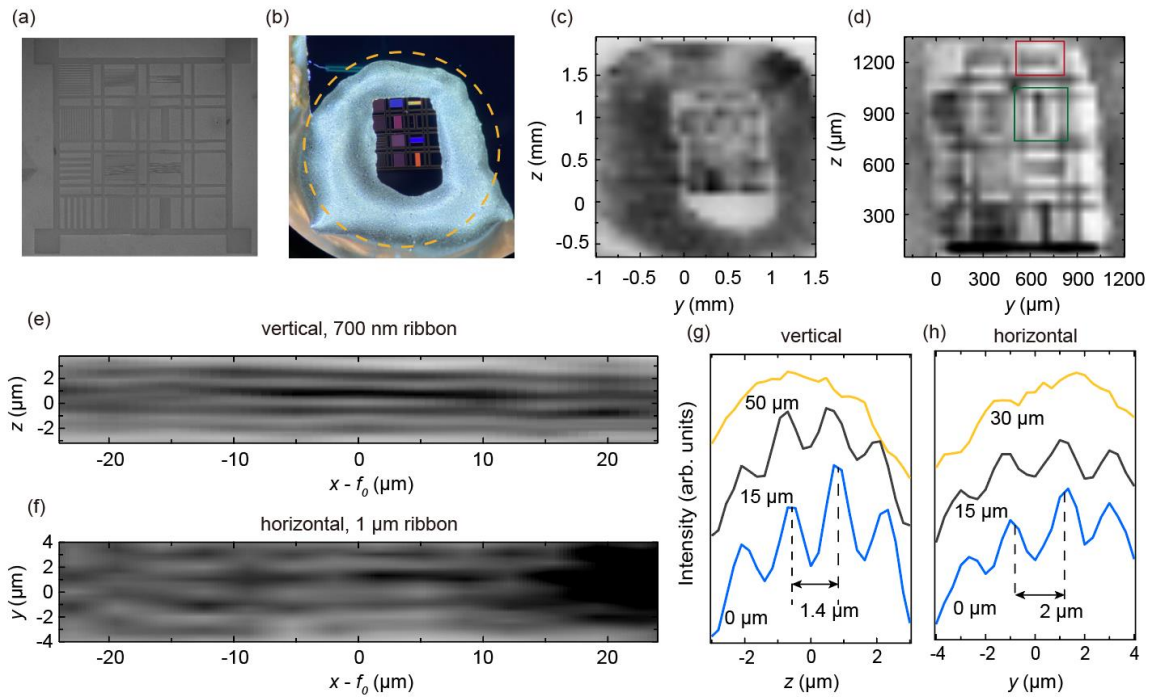

FIG. 3. The scanning photoemission measurement on a gold pattern with different periodicity grown on Si substrate. (a, b) Electronic microscopy topography and photography of the gold pattern. Silver glue is pasted around the pattern for electrical contact. (c) Large-scale scanning photoemission intensity map over the whole gold pattern. (d) Photoemission map in the region of the gold pattern. The red and green rectangles mark the horizontal and vertical regions used for the measurements in (e) and (f). (e, f) Scanning photoemission measurement across the vertical (700 nm gold ribbon) and horizontal (1 μm gold ribbon) patterns respectively. The focusing lens scans along the beam path to reach the focal point. (g, h) Line profiles of the data in (e) and (f) at different distances from the focal point.

Next, we perform scanning photoemission measurement on a microscale gold pattern deposited on a Si substrate. As shown in Fig. 3(a), the pattern is composed of horizontal and vertical periodic gold ribbons. The widths of the gold ribbons range from 20 μm to 0.7 μm, as characterized by the scanning electron microscope (SEM) topography in Fig. 3(a). Since the gold ribbons are deposited on the silicon wafer with $SiO_2$ on top, the photoemission intensity shows drastic variation across the pattern. Figure 3(b) shows the photography of the device, around which we use silver glue for electrical contact to prevent charging effect. Figure 3(c) shows the 2D real-space scanning photoemission measurement in a 2.5 × 2.7 $mm^2$ scale, where the silver glue and the patterns can be clearly distinguished. By zoom-in measurement in the pattern region, the pattern structure with different periodicities can be resolved [Fig. 3(d)]. After identifying the pattern with specific periodicity, we can accurately shift the targeting region [red and green rectangle in Fig. 3(d)] to the focused beam and measure the local photoelectrons. Figures 3(e) and (f) show the focusing scan results across the vertical 700 nm [Fig. 3(e)] and horizontal 1 μm ribbons [Fig. 3(f)]. Clearly, around the focal point, the sub-micron structures can be well resolved as shown by the line profiles in Figs. 3(g) and (h). In addition, it is found that the vertical 700 nm ribbon can be well resolved within the focal depth of about 50 μm [Figs. 2(e) and 3(e)]. It's worth noting that there is an angle of about 30° between the beam path and sample surface in the horizontal direction. Therefore, the spot projection and gold ribbon (with the thickness of about 100 nm) shadow will influence the characterization of spatial resolution in the horizontal direction. All experiments above confirm that our LMS-ARPES indeed establishes sub-micron spatial resolution.

## IV. SPIN-POLARIZED MEASUREMENTS

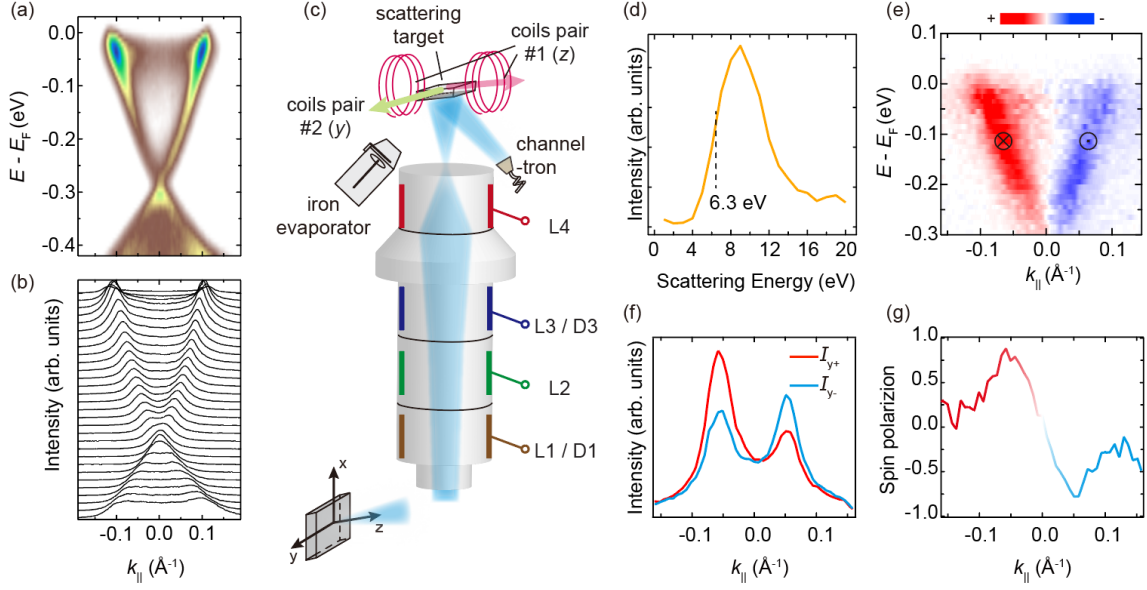

FIG. 4. Spin-ARPES measurements on Bi$_2$Se$_3$. (a) ARPES intensity map of Bi$_2$Se$_3$. (b) The corresponding momentum distribution curves (MDCs). (c) Schematic diagram of the spin detection process of the VLEED detector. (d) Scattering curve of the surface passivated scattering target. (e) y-component of spin-resolved band dispersion of Bi$_2$Se$_3$. (f) MDCs of the spin-polarized ARPES spectra integrated in an energy window of 20 meV around 0.145 eV below $E_F$. (g) The MDC of the differential map in (e) integrated in an energy window of 20 meV near 0.145 eV below $E_F$.

Figure 4 shows the spin-resolved ARPES measurements on the prototypical topological insulator Bi$_2$Se$_3$. The topological surface states (TSSs) show a gapless Dirac cone near 270 meV below $E_F$, which can be resolved in ARPES mapping [Fig. 4(a)] and corresponding momentum distribution cures (MDCs) [Fig. 4(b)], in good agreement with the previous results [51]. It is noteworthy that we do not observe strong space charge effect in the ARPES measurement.

Our LMS-ARPES system is equipped with a two-dimension VLEED detector, enabling the measurement of spin polarizations along two directions [sample normal $z$ and in-plane $y$ that is perpendicular to the analyzer slit, see Fig. 4(c)] by polarizing the scattering target using magnetic coils. 3D spin-polarized measurement can be achieved by installing another VLEED detector in the system, which is a future plan.

Figure 4(c) shows the schematic diagram of the spin detection process of the VLEED detector. High-quality iron film is prepared *in-situ* as the scattering target. The W(100) crystal is firstly cleaned by oxidized annealing; then the iron film with thickness of about 10 nm is evaporated on the substrate followed by oxidization process to passivate the target surface so that the lifetime of the scattering target can be significantly prolonged [49].

In the spin-polarized measurement, the photoelectrons first pass through the semi-sphere electron analyzer where their energy and emission angle are recorded. Then they are deflected through a hole in the analyzer detector and guided into the VLEED chamber. Several groups of electrostatic voltages [L1~L4 in Fig. 4(c)] are used in the VLEED chamber to rectify and send the photoelectron beam to scatter at magnetically polarized target. After being reflected by the scattering target, the spin-polarized electrons are finally collected by the channeltron. Figure 4(d) shows the scattering curve [intensity of spin-polarized electrons as a function of scattering energy (SE)], where the position of the maximum rising slope indicates the scattering energy for the highest spin detection efficiency (around SE = 6.3 eV) [52]. Near the peak of the scattering curve (~ 8 eV), the reflection of the scattering target (the ratio between incident and reflected electrons) is higher than 20% (table I), which guarantees the high-efficiency of spin-polarized measurements. Two pairs of electric coils are installed to polarize the scattering target along the *y* or *z* direction [the coils pair #2 are not shown in Fig. 4(c) for clarity]. When the scattering target is polarized by the coils, the scattered electron intensity shows a high asymmetry depending on the relative orientation of electron-spins $\sigma$ and the magnetization $\mu$ of the scattering target. Therefore, the spin polarization along *y* and *z* can be measured.

To test the spin resolution of our LMS-ARPES system, we perform spin-resolved ARPES measurement on prototypical topological insulator $Bi_2Se_3$. Figure 4(e) shows the differential map of the *y*-component (in-plane tangential direction of the Fermi surface) of the spin-polarization of the TSSs. Clearly the left and right branches of the Dirac cone show opposite spin-polarization along the *y* direction, confirming the spin-momentum locking of the TSSs [53]. Figure 4(f) shows the

MDCs of the raw intensity map of photoelectrons with different spin polarization, which clearly shows a spin-dependent asymmetry. Figure 4(g) shows the spin-polarization rate as a function of electron momentum, $P = A/S$, where $A = \frac{I_+ - I_-}{I_+ + I_-}$ is the asymmetry of the spin-polarized ARPES spectra and $S$ is the Sherman function. Near $k_\parallel$ = -0.065 Å$^{-1}$, the maximum asymmetry reaches 23%. Considering the value of Sherman function $S$ at scattering energy of 6.3 eV, $S = 0.275$ [54, 55], we obtain the maximum spin polarization rate of the TSSs along $y$ of about 84% [Fig. 4(g)], consistent with previous studies [44, 45], suggesting the high spin detection efficiency of our LMS-ARPES.

## V. CONCLUSIONS

We have developed a 177 nm laser-based ARPES system with sub-micron spatial resolution and high-efficiency spin detection. The spatial resolution of our LMS-ARPES is characterized by different methods to be less than 1 μm. Spin-polarization rate as high as 84% was obtained in the TSSs of Bi$_2$Se$_3$, confirming its spin-momentum locking property. The high spatial resolution and efficient spin detection of the LMS-ARPES system enable the investigation of local fine electronic structure and spin structure of novel quantum materials, which will serve as a powerful tool in the exploration of condensed matter physics.

## DATA AVAILABILITY

The data supporting the advances of this work are available from the corresponding authors.

## ACKNOWLEDGMENTS

This work is funded by the National Natural Science Foundation of China (Grants No. 11427903 and No. 11774190).

## Reference

[1] R. M. Martin, *Electronic structure: basic theory and practical methods* (Cambridge university press, 2020).
[2] P. D. King, S. Picozzi, R. G. Egdell and G. Panaccione Chem. Rev. **121**, 2816 (2020).

[3]S. Hüfner, *Very high resolution photoelectron spectroscopy* (Springer, 2007).

[4]A. Damascelli, Z. Hussain and Z.-X. Shen Rev. Mod. Phys. **75**, 473 (2003).

[5]J. A. Sobota, Y. He and Z.-X. Shen Rev. Mod. Phys. **93**, 025006 (2021).

[6]D. Lu, I. M. Vishik, M. Yi, Y. Chen, R. G. Moore and Z.-X. Shen Annu. Rev. Condens. Matter Phys. **3**, 1 (2012).

[7]X. Zhou, W.-S. Lee, M. Imada, N. Trivedi, P. Phillips, H.-Y. Kee, P. Törmä and M. Eremets Nature Reviews Physics **3**, 462 (2021).

[8]J. Denlinger, G.-H. Gweon, J. Allen, C. Olson, Y. Dalichaouch, B.-W. Lee, M. Maple, Z. Fisk, P. Canfield and P. Armstrong Physica B: Condensed Matter **281**, 716 (2000).

[9]Z. Yin, X. Du, W. Cao, J. Jiang, C. Chen, S. Duan, J. Zhou, X. Gu, R. Xu and Q. Zhang Phys. Rev. B **105**, 245106 (2022).

[10]L. Kang, X. Du, J. Zhou, X. Gu, Y. Chen, R. Xu, Q. Zhang, S. Sun, Z. Yin and Y. Li Nat. Commun. **12**, 1 (2021).

[11]Y. Li, J. Jiang, H. Yang, D. Prabhakaran, Z. Liu, L. Yang and Y. Chen Phys. Rev. B **97**, 115118 (2018).

[12]L. Yang, Y. Zhang, H. Ou, J. Zhao, D. Shen, B. Zhou, J. Wei, F. Chen, M. Xu and C. He Phys. Rev. Lett. **102**, 107002 (2009).

[13]F. Bisti, V. A. Rogalev, M. Karolak, S. Paul, A. Gupta, T. Schmitt, G. Güntherodt, V. Eyert, G. Sangiovanni and G. Profeta Phys. Rev. X **7**, 041067 (2017).

[14]X. Xu, Y. Li, S. Duan, S. Zhang, Y. Chen, L. Kang, A. Liang, C. Chen, W. Xia and Y. Xu Phys. Rev. B **101**, 201104 (2020).

[15]Y. Chen, X. Gu, Y. Li, X. Du, L. Yang and Y. Chen Matter **3**, 1114 (2020).

[16]M. Z. Hasan, G. Chang, I. Belopolski, G. Bian, S.-Y. Xu and J.-X. Yin Nat. Rev. Mater. **6**, 784 (2021).

[17]B. Lv, T. Qian and H. Ding Rev. Mod. Phys. **93**, 025002 (2021).

[18]C. Bao, P. Tang, D. Sun and S. Zhou Nature Reviews Physics **4**, 33 (2022).

[19]C. Bao, H. Zhong, S. Zhou, R. Feng, Y. Wang and S. Zhou Rev. Sci. Instrum. **93**, 013902 (2022).

[20]G. Liu, G. Wang, Y. Zhu, H. Zhang, G. Zhang, X. Wang, Y. Zhou, W. Zhang, H. Liu and L. Zhao Rev. Sci. Instrum. **79**, 023105 (2008).

[21]C. Yan, E. Green, R. Fukumori, N. Protic, S. H. Lee, S. Fernandez-Mulligan, R. Raja, R. Erdakos, Z. Mao and S. Yang Rev. Sci. Instrum. **92**, 113907 (2021).

[22]T. Kiss, T. Shimojima, K. Ishizaka, A. Chainani, T. Togashi, T. Kanai, X.-Y. Wang, C.-T. Chen, S. Watanabe and S. Shin Rev. Sci. Instrum. **79**, 023106 (2008).

[23]M. Kitamura, S. Souma, A. Honma, D. Wakabayashi, H. Tanaka, A. Toyoshima, K. Amemiya, T. Kawakami, K. Sugawara and K. Nakayama Rev. Sci. Instrum. **93**, 033906 (2022).

[24]G. Rohde, A. Hendel, A. Stange, K. Hanff, L.-P. Oloff, L. Yang, K. Rossnagel and M. Bauer Rev. Sci. Instrum. **87**, 103102 (2016).

[25]J. Koralek, J. Douglas, N. Plumb, J. Griffith, S. Cundiff, H. Kapteyn, M. Murnane and D. Dessau Rev. Sci. Instrum. **78**, 053905 (2007).

[26]M. Hoesch, T. Kim, P. Dudin, H. Wang, S. Scott, P. Harris, S. Patel, M. Matthews, D. Hawkins and S. Alcock Rev. Sci. Instrum. **88**, 013106 (2017).

[27]Y. He, I. M. Vishik, M. Yi, S. Yang, Z. Liu, J. J. Lee, S. Chen, S. N. Rebec, D. Leuenberger and A. Zong Rev. Sci. Instrum. **87**, 011301 (2016).

[28]S. C. Speller, P. Dudin, S. Fitzgerald, G. M. Hughes, K. Kruska, T. B. Britton, A. Krzton-Maziopa, E. Pomjakushina, K. Conder, A. Barinov and C. R. M. Grovenor Phys. Rev. B **90**, 024520 (2014).

[29]M. Bendele, A. Barinov, B. Joseph, D. Innocenti, A. Iadecola, A. Bianconi, H. Takeya, Y. Mizuguchi, Y. Takano, T. Noji, T. Hatakeda, Y. Koike, M. Horio, A. Fujimori, D. Ootsuki, T. Mizokawa and N. L. Saini Sci. Rep. **4**, 5592 (2014).

[30]M. Cattelan and N. A. Fox Nanomaterials **8**, 284 (2018).

[31]P. V. Nguyen, N. C. Teutsch, N. P. Wilson, J. Kahn, X. Xia, A. J. Graham, V. Kandyba, A. Giampietri, A. Barinov and G. C. Constantinescu Nature **572**, 220 (2019).

[32]M. Utama, R. J. Koch, K. Lee, N. Leconte, H. Li, S. Zhao, L. Jiang, J. Zhu, K. Watanabe and T. Taniguchi Nat. Phys. **17**, 184 (2021).

[33]D. Pei, B. Wang, Z. Zhou, Z. He, L. An, S. He, C. Chen, Y. Li, L. Wei and A. Liang arXiv preprint arXiv:2205.13788, (2022).

[34]R. Noguchi, M. Kobayashi, Z. Jiang, K. Kuroda, T. Takahashi, Z. Xu, D. Lee, M. Hirayama, M. Ochi and T. Shirasawa Nat. Mater. **20**, 473 (2021).

[35]R. J. Koch, C. Jozwiak, A. Bostwick, B. Stripe, M. Cordier, Z. Hussain, W. Yun and E. Rotenberg, (Taylor & Francis, 2018).

[36]J. Avila, A. Boury, B. Caja-Muñoz, C. Chen, S. Lorcy and M. C. Asensio, *Proceedings of Journal of Physics: Conference Series,* (IOP Publishing).

[37]P. Dudin, P. Lacovig, C. Fava, E. Nicolini, A. Bianco, G. Cautero and A. Barinov Journal of synchrotron radiation **17**, 445 (2010).

[38]H. Iwasawa, P. Dudin, K. Inui, T. Masui, T. K. Kim, C. Cacho and M. Hoesch Phys. Rev. B **99**, 140510 (2019).

[39]E. Rotenberg and A. Bostwick Journal of Synchrotron Radiation **21**, 1048 (2014).

[40]Y. Mao, D. Zhao, S. Yan, H. Zhang, J. Li, K. Han, X. Xu, C. Guo, L. Yang and C. Zhang Light: Science & Applications **10**, 1 (2021).

[41]K. Gotlieb, Z. Hussain, A. Bostwick, A. Lanzara and C. Jozwiak Rev. Sci. Instrum. **84**, 093904 (2013).

[42]F. Ji, T. Shi, M. Ye, W. Wan, Z. Liu, J. Wang, T. Xu and S. Qiao Phys. Rev. Lett. **116**, 177601 (2016).

[43]C. Jozwiak, J. Graf, G. Lebedev, N. Andresen, A. Schmid, A. Fedorov, F. El Gabaly, W. Wan, A. Lanzara and Z. Hussain Rev. Sci. Instrum. **81**, 053904 (2010).

[44]C. Jozwiak, C.-H. Park, K. Gotlieb, C. Hwang, D.-H. Lee, S. G. Louie, J. D. Denlinger, C. R. Rotundu, R. J. Birgeneau and Z. Hussain Nat. Phys. **9**, 293 (2013).

[45]Z.-H. Zhu, C. Veenstra, S. Zhdanovich, M. Schneider, T. Okuda, K. Miyamoto, S.-Y. Zhu, H. Namatame, M. Taniguchi and M. Haverkort Phys. Rev. Lett. **112**, 076802 (2014).

[46]M. Erbudak and N. Müller Appl. Phys. Lett. **38**, 575 (1981).

[47]E. D. Schaefer, S. Borek, J. Braun, J. Minár, H. Ebert, K. Medjanik, D. Kutnyakhov, G. Schönhense and H.-J. Elmers Phys. Rev. B **95**, 104423 (2017).

[48]D. Tillmann, R. Thiel and E. Kisker Zeitschrift für Physik B Condensed Matter **77**, 1 (1989).

[49]R. Bertacco, M. Merano and F. Ciccacci Appl. Phys. Lett. **72**, 2050 (1998).

[50]A. H. Firester, M. Heller and P. Sheng Appl. Opt. **16**, 1971 (1977).

[51]Y. Xia, D. Qian, D. Hsieh, L. Wray, A. Pal, H. Lin, A. Bansil, D. Grauer, Y. S. Hor and R. J. Cava Nat. Phys. **5**, 398 (2009).

[52]T. Okuda, Y. Takeichi, Y. Maeda, A. Harasawa, I. Matsuda, T. Kinoshita and A. Kakizaki Rev. Sci. Instrum. **79**, 123117 (2008).

[53]C. Jozwiak, J. A. Sobota, K. Gotlieb, A. F. Kemper, C. R. Rotundu, R. J. Birgeneau, Z. Hussain, D.-H. Lee, Z.-X. Shen and A. Lanzara Nat. Commun. **7**, 1 (2016).

[54]M. Escher, N. B. Weber, M. Merkel, L. Plucinski and C. M. Schneider e-Journal of Surface Science and Nanotechnology **9**, 340 (2011).

[55]A. Winkelmann, D. Hartung, H. Engelhard, C.-T. Chiang and J. Kirschner Rev. Sci. Instrum. **79**, 083303 (2008).